\documentstyle [12pt,epsf] {article}

\newcommand{\be}{\begin{equation}}
\newcommand{\ee}{\end{equation}}
\newcommand{\beqs}{\begin{eqnarray}}
\newcommand{\eeqs}{\end{eqnarray}}

\def\({\left(}
\def\){\right)}

\def\ni{\noindent}

\def\f{\frac}

\def\bb{\bibitem}
\def\nn{\nonumber}
\def\N{${\cal N}$}

\begin{document}

\begin{titlepage}

\begin{flushright}
\begin{tabular}{l} YITP-SB-99-68 \\ hep-th/0001051 \\ Nov, 1999
\end{tabular}
\end{flushright}

\vspace{8mm}
\begin{center} {\Large Hyperk\"ahler Quotients, Mirror Symmetry, and F-theory }

\vspace{20mm}

I.Y. Park*\footnote{email: ipark@insti.physics.sunysb.edu} and
R. von Unge**\footnote{email:unge@physics.muni.cz}

\vspace{6mm} * C.N. Yang Institute for Theoretical Physics \\ 
State University of New York	\\ 
Stony Brook, N. Y. 11794-3840 \\

\vspace{4mm} ** Department of Theoretical Physics and Astrophysics\\
             Faculty of Science, Masaryk University\\
             Kotl\'{a}\v{r}sk\'{a} 2, CZ-611 37, Brno, Czech Republic

\vspace{20mm}

{\bf Abstract}
\end{center}
\indent Using the graphical method developed in hep-th/9908082, we obtain the full curve corresponding to the hyperk\"ahler quotient from the extended E$_7$ Dynkin diagram. As in the E$_6$ case discussed in the same paper above, the resulting curve is the same as the one obtained by Minahan and Nemeschansky. Our results seem to indicate that it is possible to define a
 {\em generalized Coulomb branch} such that {\em four} dimensional mirror
 symmetry would act by interchanging the generalized Coulomb branch with
 the Higgs branch of the dual theory. To understand these phenomena, we discuss mirror symmetry and F-theory compactifications probed by D3 branes.

\vspace{35mm}

\end{titlepage}
\newpage
\setcounter{page}{1}
\pagestyle{plain}
\pagenumbering{arabic}
\renewcommand{\thefootnote}{\arabic{footnote}} \setcounter{footnote}{0}

\section{Introduction}

\indent String theory provides a very powerful setting for the study
of gauge theories. Gauge theories can be constructed through the
geometrical engineering \cite{KKV}  by compactifying string theory on
Calabi-Yau manifolds with appropriate Hodge numbers and singularities.
They can also be realized as the world volume theories on extended
objects such as D-branes \cite{Pol}. The authors of \cite{DM1} studied
the gauge theories obtained by placing D-branes on orbifold
singularities and introduced `quiver diagrams', which summarize the
gauge group structures and matter contents of the gauge theories. They
considered D-5 branes and noticed that the moduli space of D-brane
ground states is a ALE space described by a hyperk\"ahler quotient. 
         
The hyperk\"ahler quotient was introduced in \cite{LR}; it was
mathematically refined in \cite{HKLR}. One way to construct it is to
gauge isometries of a non-linear sigma model in such a way as to
preserve \N=2 supersymmetry. In the spirit of \cite{Seiberg}, a
graphical method was invented in \cite{LRV} and used to obtain the
curve that corresponds to a hyperk\"ahler quotient of a linear
space. In particular, it was applied to the hyperk\"ahler quotients
constructed from the extended Dynkin diagrams of A$_k$, D$_k$ series
and E$_6$ case.  

Remarkably, the Higgs branch of a quiver gauge theory based on the
extended E$_6$ Dynkin diagram turned out to be identical, when it was
expressed in terms of E$_6$ Casimir invariants, to the
curve\footnote{We will call this curve {\em the generalized Coulomb
branch}.} with E$_6$ global symmetry obtained by Minahan and 
Nemeschansky some time ago \cite{MN} and
 later by the authors of \cite{NTY}. In this article, we work out the
 full\footnote{The orbifold limits of E$_7$ and E$_8$ (and some other
 higher order quiver diagrams) were considered in \cite{LRV}.} curve
 corresponding to the E$_7$ extended Dynkin diagram. The resulting
 curve is again equal to the generalized Coulomb branch with E$_7$
 global symmetry computed in \cite{MN2} and \cite{NTY}\footnote{ In
 E$_7$ case, it is easier to compare with \cite{NTY} since the authors
 used E$_7$ Casimir invariants, while the authors of \cite{MN2}
 expressed their curve in terms of the SO(12)$\times$SU(2) Casimir
 invariants, as we discussed in section 2.}.  
              
We understand the origin of these phenomena through mirror symmetry and F-theory \cite{V} compactifications \cite{Sen,DM2}. D3 branes are used to probe the singularities of the backgrounds under consideration \cite{BDS}. The relevant F-theory compactifications for our purpose are the ones which give rise to E$_7$ gauge group. The E$_7$ {\em global} symmetry is realized on the world volume theory of the D3 branes. 

It is the physics near such singularities that is responsible for 
the field theory limit of string compactifications \cite{KKLMV}. The 
mirror geometry of ADE singularities was discussed \cite{KMV} in the 
context of type II strings. On the dual backgrounds, the gauge groups 
of the dual superconformal field theories are given by a product of 
U(n$_i$) groups. The n$_i$'s are given by multiples of the Dynkin numbers 
of the nodes in the corresponding Dynkin diagrams.

Mirror symmetry is well understood in three dimensions \cite{IS,HOV}
where both the Higgs branch and the Coulomb branch are hyperk\"ahler
manifolds. They get interchanged under the action of mirror
symmetry. In the four dimensional models we consider in this paper,
the Coulomb branch of the original theory is a Riemann surface, which
is real two dimensional, whereas the Higgs branch of the dual theory
has real four dimensions. What we find in this paper seems to 
indicate that mirror symmetry in
these four dimensional models acts in such a way that it is the {\em
generalized} Coulomb branch (rather than the Coulomb branch) of the
original gauge theory that gets interchanged with the Higgs branch of
the mirror dual theory.\footnote{There is a natural relation between
the generalized Coulomb branch in four dimensions and the Coulomb
branch in three dimensions, as discussed in section 3.} 

More intuitive understanding of the origin of the identity between the
curve we compute and the generalized Coulomb branch seems possible by
applying various string dualities to the system under
consideration. We briefly discuss this point in section 3 with a heuristic
example using the D7-D3 brane system.

The organization is as follows. After briefly reviewing the
hyperk\"ahler quotients, we present the calculation of E$_7$ case in
section 2. The final form of the curve is given in Appendix A. It is
expressed in terms of E$_7$ Casimirs, $P_i$, whose definition is given
in Appendix A. As in the case of E$_6$, the curve obtained is the
generalized Coulomb branch with E$_7$ global symmetry. The generalized
Coulomb branch can also be expressed in terms of E$_7$ Casimir
invariants which we also denote as $P_i$. However, the $P_i$'s of our
curve are functions of Fayet-Iliopoulos (FI) parameters, b$_j$, while
the $P_i$'s of the generalized Coulomb branch are functions of mass
parameters\footnote{They are associated with relevant deformations of
the superconformal field theory under consideration.}, m$_k$. In
anticipation of mirror symmetry, we find the relations between b's and
m's which render $P_i(b)=P_i(m)$.\footnote{These relations reflect the
fact that under mirror symmetry FI and mass parameters get
interchanged.} In section 3, we discuss mirror symmetry and F-theory
compactifications probed by D3 branes. Section 4 includes summary and
open problems.

\section{The Hyperk\"ahler Quotient For E$_7$ Case}       
 We begin by briefly reviewing the hyperk\"ahler quotients and refer
 the reader to \cite{LR,HKLR,LRV} (and the references therein) for
 more details. Intuitively speaking the (hyper)k\"{a}hler quotient is a
 method that, starting with a space with a metric that has some
 isometry, finds a hypersurface orthogonal to the isometry direction
 and the induced metric on that surface. The method used in this paper
 will allow us to explicitly find this hypersurface as a two
 dimensional complex space embedded in $C^3$. It will also allow us to
 find the explicit dependence of this space on the Fayet-Iliopoulos
 parameters of the gauge theory we start with. In general, turning
 them on we deform the hypersurface so that it becomes non-singular.
To also find the induced metric in the hyperk\"{a}hler quotient one should
 introduce a non-linear sigma model with the original space as its
 target space. Gauging the isometries of this non-linear sigma model while
 preserving \N=2 susy gives rise to the hyperk\"ahler quotient. More
 specifically, consider a sigma model with isometries. To elevate the
 isometries to local symmetries, introduce an \N=2 vector multiplet,
 which consists of an \N=1 vector multiplet and an \N=1 chiral
 multiplet, denoted respectively as $V$, $S$ in \cite{LR}. In \N=1
 superspace, one integrates out $V$ and $S$ by their field
 equations. Inserting the solution for $V$ field equations into the
 gauged Lagrangian and keeping the $S$ field equations as constraints
 gives the K\"ahler potential of the quotient space. The constraints
 from $S$ field equations can be represented graphically and are given
 in figure 1\footnote{ Figure 1 and Figure 3 are taken from
 \cite{LRV}.}. The gauge groups and the representations appropriate
 for the the construction of ALE spaces are summarized by the extended
 Dynkin diagrams \cite{K}, as in Figure 2.   

\begin{figure}
{\epsfxsize=16cm\epsfbox{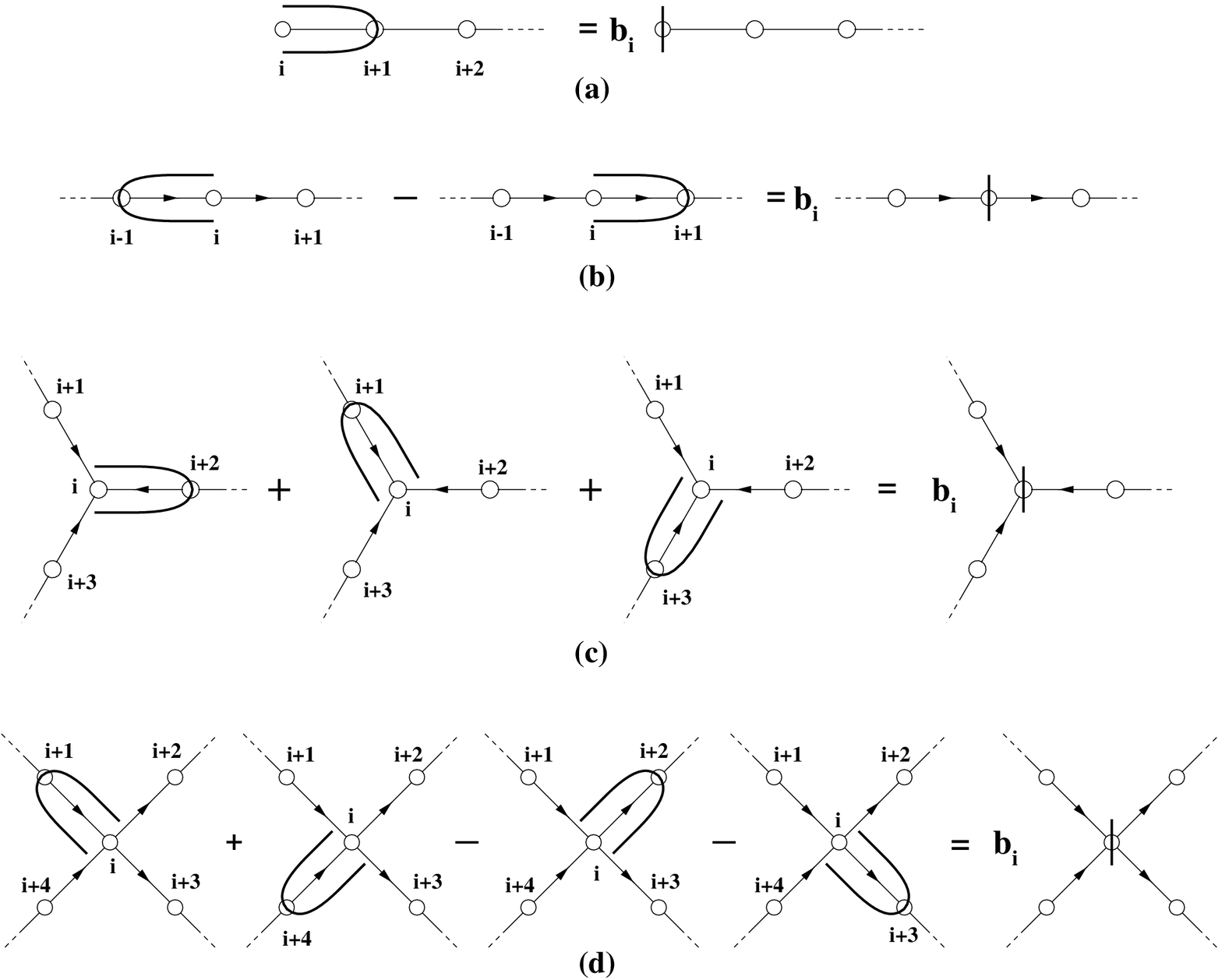}}
\noindent{\bf Figure 1}: The bug calculus. $b_i$ is the Fayet-Iliopoulos
parameter associated to the $i$'th node, and a vertical bar through the
$i$'th node represents a $U(N_i)$ Kronecker-$\delta$.  
\end{figure}             

\begin{figure}
{\epsfxsize=16cm\epsfbox{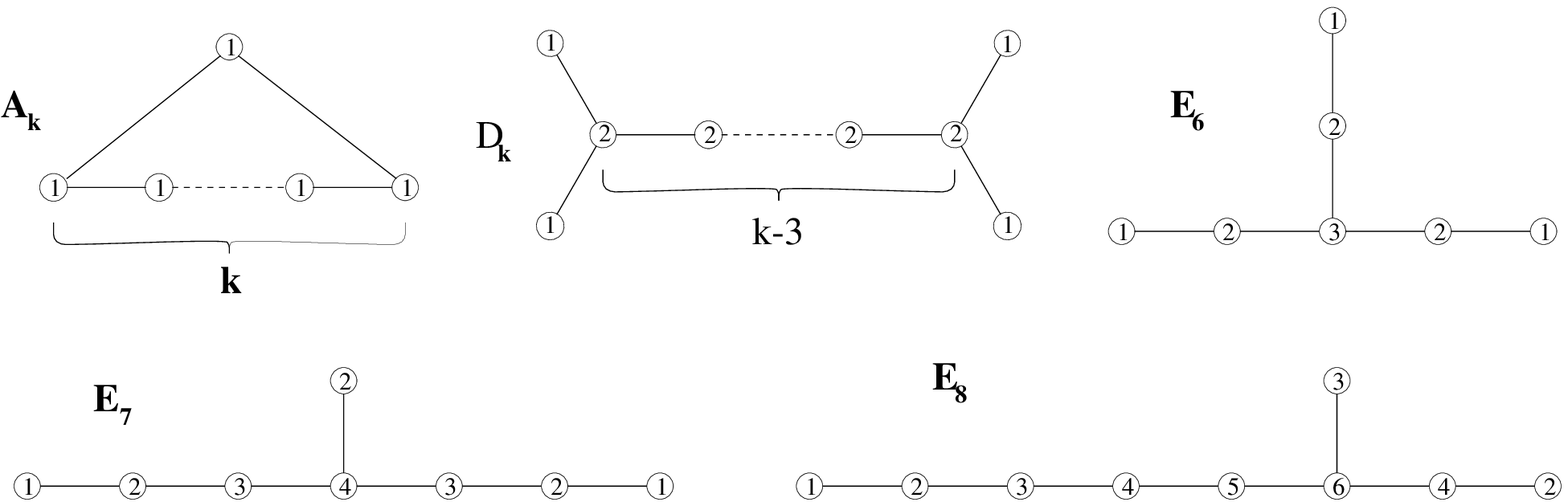}}
\centerline{\bf Figure 2}   
\end{figure}

\begin{figure}
{\epsfxsize=16cm\epsfbox{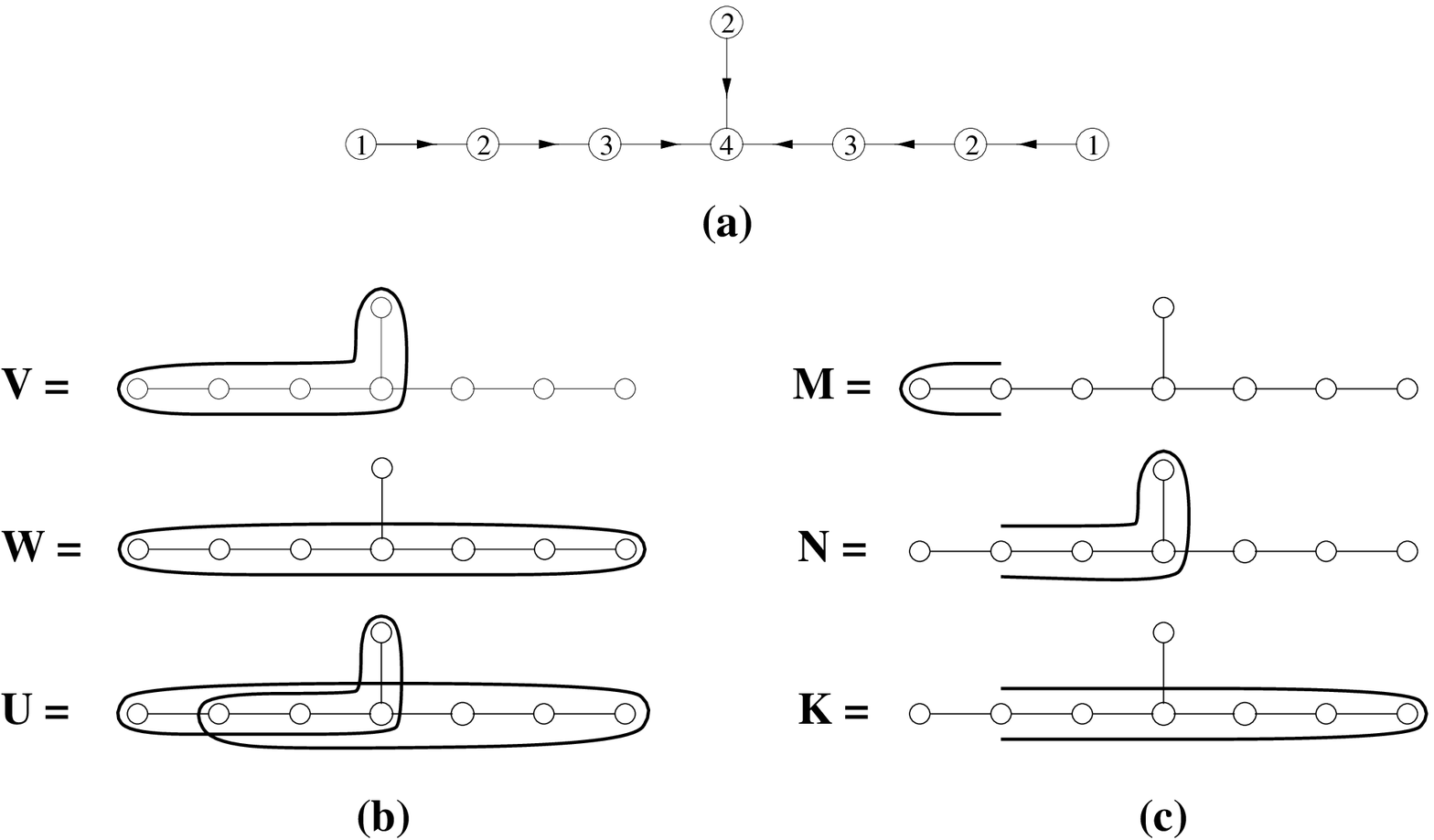}}
\centerline{{\bf Figure 3}: The $E_7$ invariants and some useful
matrices.}
\end{figure}     

Now, we compute the hyperk\"ahler quotient corresponding to the E$_7$
extended Dynkin Diagram given in Figure 3(a). In addition to the
Dynkin numbers in the same figure, we label the nodes by assigning 1
to the far left node and 2 to the next one, etc. The upper node in the
middle is referred to as the eighth node. We closely follow \cite{LRV}
with a convenient set of variables defined in Figure 3(b).  

Consider the highest order invariant, $U$, and its orientation
reversed diagram $\bar{U}$. The product of these two diagrams can be
written as\footnote{In the orbifold limit, $U=-\bar{U}$, but this is
not true in the presence of the Fayet-Iliopoulos terms.}        
\be
U\bar{U}=W\,Tr(MNKN) \label{start}
\ee

\ni One can use the so called Schouten identity to rewrite $\bar{U}$
in terms of the variables defined in Figure 3(b): The relevant
Schouten identity is 
\beqs
 Tr (\{M,N\} K) = Tr (MN) Tr (K)  + Tr(MK) Tr (N)
                  +Tr (NK) Tr (M) - Tr (M) Tr (N) Tr (K)\nn\\ \label{Schouten}
\eeqs

\ni Noting the following relations, 

\beqs
Tr(\{M,N\}K) &=& U+\bar{U} \nn\\
Tr(M)  & = & b_1 \nn\\
Tr(MN) & = & V \nn\\
Tr(MK) & = & W  \label{rel}
\eeqs

\ni one obtains,

\be
\bar{U} = -U+Tr(K)\;V+Tr(N)\;W+b_1 Tr(NK)-b_1Tr(N)Tr(K)
                         \;\; \nn\\ \label{Ubar} 
\ee
where $b_1$ is the FI term associated with the first node of the
Dynkin diagram. To rewrite the right hand side of (\ref{start}),
consider 

\beqs
  0     & = &  N^{[ k_{2}}_{k_{1}}K^{k_{3}}_{k_{2}} 
                         N^{k_{4}]}_{k_{3}}M^{k_1}_{k_4} \nn\\
        & = & Tr(MNKN)+Tr(NNKM)-Tr(NNM)Tr(K)-Tr(NKM)Tr(N)\nn\\
        &   &   +Tr(NM)Tr(K)Tr(N)-Tr(NM)Tr(KN) \nn\\ 
\eeqs         

\ni Applying the Schouten identity (\ref{Schouten}) again to
$Tr(NNKN)$ and $Tr(NNM)$ leads to 

\beqs
Tr(MNKN) &=& Tr(MN)Tr(NK)-\f{1}{2}Tr(MK)Tr(NN)+\f{1}{2}Tr(MK)Tr(N)^2\nn\\
         & & +\f{1}{2}Tr(NN)Tr(M)Tr(K)-\f{1}{2}Tr(M)Tr(K)Tr(N)^2 \nn\\  
         &=&     Tr(NK)\;V+\left[-\f{1}{2}Tr(NN)+Tr(N)^2\right] \;W \nn\\
         & & \;\;\;\;\;\;\;\;\;\;\;\;\;\;\;\;\;\;\;\;\;\;\;\;\;\;\;\;\;\;
                     +\f{b_1}{2}Tr(K)Tr(NN)
                      -\f{b_1}{2}Tr(K)Tr(N)^2     \label{RHS}                  \eeqs
\ni where the second equality follows from (\ref{rel}). Substituting
(\ref{Ubar}) and (\ref{RHS}) into (\ref{start}) gives 

\beqs
& &  U\left(-U+Tr(K)\;V+Tr(N)\;W+b_1 Tr(NK)-b_1Tr(N)Tr(K)
                         \;\;\right) \nn\\                   
& & = W\;\left(  Tr(NK)\;V+\left[-\f{1}{2}Tr(NN)+Tr(N)^2\right] \;W
                                     +\f{b_1}{2}Tr(K)Tr(NN)
                      -\f{b_1}{2}Tr(K)Tr(N)^2   \right) \nn\\ \label{curve1}
\eeqs
       
Therefore the whole task of finding the curve is reduced to the
computation of $Tr(N), Tr(K), \\ 
Tr(NN)$ and $Tr(NK)$. It is a simple exercise to compute $T(N)$: It is 
expressed purely in terms of $b_i$'s. The other three quantities are 
more complicated to obtain: The final forms are,

\beqs
Tr(K)   &=& k_v(b_i)V+k   \nn\\
Tr(NN)  &=& 2W+n_v(b_i)\;V+n(b_i)  \nn\\
Tr(NK)  &=& m_w(b_i)W-V^2+m_v(b_i)V+m(b_i) \label{NK} 
\eeqs
where the coefficients are functions of the FI parameters, $b_i$, as
indicated. Since these coefficients are lengthy, we will not present
them explicitly. However, once we make the change of variables
discussed below, the coefficients can be expressed in terms of E$_7$
Casimir invariants. This makes the curve simple enough to
present. Upon substitution of (\ref{NK}) into (\ref{curve1}), we
obtain, 

\beqs
& & U^2-W^3-V^3W   \nn\\
& & + U\;(\;(-b_1m_w-l)\;W+(b_1-k_v)\;V^2+(-b_1m_v+b_1lk_v-k)\;V
      +b_1(lk-m)\;) \nn\\
& & + W^2\left( \left[m_w-\f{1}{2}n_v+b_1k_v\right]V
                     +\f{1}{2}l^2+b_1k-\f{n}{2}\right) \nn\\
& & + W\left( \left[\f{b_1}{2}n_vk_v+m_v \right]V^2+
                \left[m+\f{b_1}{2}nk_v+\f{b_1}{2}n_vk
                     -\f{b_1}{2}l^2k_v  \right]V
                     -  \f{b_1}{2}l^2k+\f{b_1}{2}nk\right)=0
                                     \label{curve2}    \nn\\
\eeqs
where $l\equiv Tr(N)$. To put this curve into the standard form, we
perform the following change of variables, 

\beqs
U &=& X-\f{1}{2}[\; (b_1m_w-l)W+(b_1-k_v)V^2+(-b_1m_v+b_1lk_v-k)V
                         + b_1(lk-m)\;]   \nn\\
V &=&  \left. Z+  \f{1}{6}{b_1}^2 m_w+\f{1}{18}b_1k_vn_v \right.\nn \\           & &   \;\;\;\;\;\;\;\;\;\;\;  + \f{1}{18}b_1k_vm_w+\f{1}{3}m_v
               -\f{1}{9}n_vm_w+\f{1}{36}n_v^2 -\f{1}{6}k_vl 
           +\f{1}{6}b_1l+\f{1}{9}b_1^2k_v^2+\f{1}{9}m_w^2  \nn\\   
W &=& Y+ \left[-\f{1}{6}n_v+\f{1}{3}m_w+\f{1}{3}b_1k_v \right] Z 
                      +\f{1}{3}b_1k-\f{1}{12}b_1^2m_w^2-\f{1}{6}b_1lm_w
                         +\f{1}{12}l^2-\f{1}{6}n \nn \\
  & &  +\f{1}{6}\left[2b_1k_v-n_v+2m_w \right] 
                \left[ \f{1}{6}{b_1}^2 m_w
                          +\f{1}{18}b_1k_vn_v+\f{1}{18}b_1k_vm_w
                                    +\f{1}{3}m_v \right. \nn\\
  & & \;\;\;\;\;\;\;\;\;\;\;\;\;\;\;\;\;\;\;\;\;\;\;\;\;\;\;\;\;\;\;\;\;\;\;
        \; 
               \left. -\f{1}{9}n_vm_w+\f{1}{36}n_v^2 -\f{1}{6}k_vl 
           +\f{1}{6}b_1l+\f{1}{9}b_1^2k_v^2+\f{1}{9}m_w^2 \right]  \nn\\
      \label{shift}
\eeqs

\ni In terms of the new variables, $X,Y$ and $Z$, the curve becomes

\beqs
X^2 &=& Y^3+ f(Z)Y+g(Z) \nn \label{ourcurve} \\
\eeqs
where
\beqs
f(Z)&=& Z^3+\alpha_1(b_i) Z+{\alpha}_0(b_i)  \nn\\
g(Z)&=& {\beta}_4(b_i) Z^4+\beta_3(b_i) Z^3+\beta_2(b_i) Z^2
              +\beta_1(b_i) Z+\beta_0(b_i)  \nn\\  \label{curve3}
\eeqs

\ni The coefficients $\alpha$ and $\beta$ are expressible in terms of
E$_7$ Casimir invariants and are given in the Appendix A.  

We discuss the comparison of (\ref{ourcurve}) with the curve of
\cite{MN2} in Appendix B in more detail. Here we only present the
relations\footnote{It is easier to compare to \cite{NTY} since they
also use E$_7$ Casimirs. It is straightforward to check that our curve
is equal to that of \cite{NTY} if we identify our P$_i$ with their
P$_i$. } between b's and m's:

\beqs
  b_1 &=& \phi \nn\\   
  b_3 &=& m_5 - m_6 ,\nonumber\\
  b_4 &=& m_4 - m_5 ,\nonumber\\
  b_5 &=& m_3 - m_4 ,\nonumber\\
  b_6 &=& m_2 - m_3 ,\\
  b_7 &=& m_1 - m_2 ,\nonumber\\
  b_8 &=& m_5 + m_6 ,\nonumber  
\eeqs  
\ni where $\phi$ is the simple root of SU(2). Upon substitution in
  (\ref{ourcurve}), we find exactly the curve of \cite{MN2}. The mass
  parameters have a group theoretical interpretation as an orthonormal
  basis for the root space.

\section{Mirror Symmetry and F-theory }

In \cite{LRV}, it was observed that the hypermultiplet moduli space of
a model constructed from the E$_6$ extended Dynkin diagram is equal to
the generalized Coulomb branch of SU(2) gauge theory with E$_6$ global
symmetry. In this letter, we have extended this observation to the
E$_7$ case. Given this remarkable correspondence it is natural to 
conjecture that there is a mirror symmetry acting on four dimensional
gauge theories analogous to the mirror symmetry acting on three
dimensional gauge theories \cite{IS}. However, the 
Coulomb branch in four dimensions is not, in general, a
Hyperk\"{a}hler manifold so it 
cannot be directly exchanged with the Higgs branch of the gauge theory
which is a Hyperk\"{a}hler manifold. Instead we conjecture that what
is exchanged with the Higgs branch is the full four dimensional
elliptically fibered space that one obtains by fibering the Seiberg-Witten
torus over the usual Coulomb branch as the
base. In other words, the Higgs branch gets interchanged with the four
dimensional space given by the equation
\beqs
  X^2 = Y^3 + f\left(Z\right)X + g\left(Z\right)
\eeqs
where $Z$ is now interpreted as the usual coordinate on the Coulomb branch
of the gauge theory. We now try to collect some evidence for this
conjecture. 

Firstly we can connect our generalized four
dimensional mirror symmetry with 
the mirror symmetry that acts on three dimensional gauge theories
\cite{IS} by performing a dimensional reduction of the four dimensional
Seiberg-Witten theory to three dimensions. Namely, in \cite{SW2} it
was shown that as soon as we compactify the four dimensional photon
becomes two scalars which coordinatize the Seiberg-Witten torus thus
making the full space spanned by the complex curve the natural object.
It should also be noted that, in the cases where we can compare, the
map between the FI parameters and the mass terms is
exactly the same as in the three dimensional case \cite{DHOO}. Our
results thus seem to be in good agreement with the results on mirror
symmetry in three dimensional gauge theories.

For a more concrete connection between our mirror dual theories we now
consider F-theory compactifications \cite{V}. The reason for this is
that it is well known that the Seiberg-Witten theories with
exceptional global symmetries can be obtained as world volume theories
of 3-branes probing certain F-theory backgrounds
\cite{Sen,DM2,J,GZ,S}. Since the F-theory background looks like a set of
7-branes at strong coupling it is likely that through a sequence of
T-dualities and S-dualities we can map this configuration to a brane
configuration for the mirror theory. For example, starting with
ordinary D3 and D7-branes, it is not difficult to imagine a sequence
of dualities which would map the F-theory configuration to a
configuration where a D5-brane wraps 2-cycles of an $A_k$
singularity. The challenge is now to extend this to the case with
exceptional groups where the original configuration is strongly
coupled and it is not clear what happens under duality.

We could also imagine constructing the Seiberg-Witten theory through
geometrical engineering. In that case we could study the how string theory
mirror symmetry acts along the lines of \cite{KMV}. It is not a priori clear
that the mirror theory obtained this way is the mirror theory proposed in
this paper but since the three dimensional mirror symmetry can be
explained in this fashion we expect a connection also in our case.   
If this picture is true we could take the viewpoint that what we have been
doing in this paper is to "solve" the superconformal field theory with E$_7$
global symmetry using the method of geometrical engineering and mirror
symmetry as outlined in \cite{KMV}.

\section{Summary and Open Problems}

We have extended the observation made in \cite{LRV} to E$_7$ case: The
curve corresponding to the hyperk\"ahler quotient based on the E$_7$
extended Dynkin diagram is equal, when it is expressed in terms of
E$_7$ Casimirs, to the generalized Coulomb branch with E$_7$ global
symmetry. The relations between FI parameters and mass parameters were
obtained. The identity of the two curves led us to conjecture that
mirror symmetry in the four dimensional field theories we considered
should act in such a way to interchange the generalized Coulomb branch
of the original theory with the Higgs branch of the dual quiver gauge
theory. For evidence, we discussed the connection of the generalized
Coulomb branches in four dimensions to the Coulomb branches of the
three dimensional theories obtained by compactifying one dimension. We
also discussed F-theory compactifications, and IIA/B mirror symmetry. 
       
What we have shown in this article is that the complex structures of
the Higgs branch and the generalized Coulomb branch are the same. To
confirm the mirror hypothesis, we also need to show that the metrics
are the same. It will be worth studying whether our conjecture is true
in more general context. It will be also interesting to consider other
quivers and study if the resulting curves can be interpreted as the
generalized Coulomb branches of higher genera of some gauge
theories. We hope to come back to these issues and others in
\cite{PV}.

\section*{Acknowledgments}
We are very grateful to Martin Ro\v{c}ek for many useful discussions.
The work of Inyong Park was supported in part by NSF grant
PHY-97-22101 and the work of Rikard von Unge was supported in part by
the Swedish Institute.

\newpage
\section*{Appendix A}  

The coefficients in (\ref{curve3}) can be expressed in terms of $E_7$
Casimir invariants, $P_i$, which appear as the coefficient of
$x^{56-i}$ in the expansion of det$(x-v\cdot H)$. One can express
$v\cdot H$ as $v\cdot H=(v\cdot \lambda_1...v\cdot
\lambda_{56})$. $\lambda$'s are the weights of the fundamental
representation. Defining\footnote{The $\chi$'s defined here with the
factor $\f{1}{2}$ in front are more convenient because the last twenty
eight weights are given by minus the first twenty eight weights as
discussed below.} $\chi_n=\f{1}{2}Tr[(v \cdot H)^n]$ , the
coefficients in (\ref{curve3}) are 

\beqs
\alpha_1 &=&
\f{1}{240}\chi_8-\f{11}{6480}\chi_6\chi_2+\f{25}{2239488}\chi_2^4
\nn\\ 
         &=& -{\frac {2405}{2239488}}\,{{ P_2}}^{4}
           +{\frac {5}{432}}\,{ P_2}\,{ P_6}
              -{\frac {1}{60}}\,{ P_8}   \nn\\
\alpha_0 &=&  -\f{1}{3240}\chi_{12}+\f{13}{81648}\chi_{10}\chi_2
               -\f{97}{3265920}\chi_8\chi_2^2+\f{19}{233280}\chi_6^2 \nn\\
         & &\;\;\;\;\;\;\;\;\;\;\;\;\;\;\;\;\;\;
                +\f{13}{16796160}\chi_6\chi_2^3
                         +\f{103}{43535646720}\chi_2^6   \nn\\ 
         &=& {\frac {63713}{60949905408}}\,{{ P_2}}^{6}
              -{\frac {179}{7838208}}\,{{ P_2}}^{3}{ P_6}
                 -{\frac {431}{408240}}\,{ P_2}\,{ P_{10}} \nn\\
         & &        +{\frac {19}{58320}}\,{{ P_2}}^{2}{ P_8}
                -{\frac {1}{5184}}\,{{ P_6}}^{2}
                         +{\frac {1}{540}}\,{ P_{12}}
                 \nn\\
\beta_4  &=&-\f{1}{36}\chi_2 = \f{1}{36}P_2 \nn\\         
\beta_3  &=&\f{1}{216}\chi_6-\f{7}{93312}\chi_2^3  
          = -\f{1}{72}P_6+\f{169}{93312}P_2^3 \nn\\
\beta_2  &=& -\f{1}{2520}\chi_{10}+\f{1}{3780}\chi_8\chi_2
                   -\f{13}{233280}\chi_6\chi_2^2
                       +\f{17}{67184640}\chi_2^5 \nn\\
         &=& {\frac {1}{504}}\,{ P_{10}}
                       -{\frac {715}{94058496}}\,{{ P_2}}^{5}
                     +{\frac {5}{27216}}\,{{ P_2}}^{2}{ P_6}
                  -{\frac {1}{1080}}\,{ P_2}\,{ P_8}
                                                   \nn\\
\beta_1  &=& \f{1}{26796}\chi_{14}-\f{479}{18604080}\chi_{12}\chi_2
                +\f{2857}{426202560}\chi_{10}\chi_2^2 \nn\\
         & &  -\f{41}{1503360}\chi_8\chi_6-\f{6893}{13638481920}\chi_8\chi_2^3
               -\f{9233}{70140764160}\chi_6\chi_2^4 \nn\\
         & &  +\f{1249}{121772160}\chi_6^2\chi_2
                           +\f{391207}{636317012459520}\chi_2^7
                                         \nn\\  
         &=&  {\frac {78346801}{127263402491904}}\,{{ P_2}}^{7}
                   -{\frac {1}{3828}}\,{ P_{14}}
                     -{\frac {96277}{4688228160}}\,{{ P_2}}^{2}{ P_{10}}
            -{\frac {1370167}{120018640896}}\,{{ P_2}}^{4}{ P_6}\nn\\
         & &       +{\frac {56233}{5357975040}}\,{{ P_2}}^{3}{ P_8}
                +{\frac {1517}{29766528}}\,{ P_2}\,{{ P_6}}^{2}
                   +{\frac {331}{3100680}}\,{ P_2}\,{ P_{12}}
                  -{\frac {91}{1378080}}\,{ P_8}\,{ P_6} \nn\\                 \beta_0  &=& -\f{1}{265464}\chi_{18}+\f{18577}{21340120032}\chi_{14}\chi_2^2
                    +\f{397}{172020672}\chi_{12}\chi_6   \nn\\
         & & -\f{37413577}{118529123834880}\chi_{12}\chi_2^3 
             +\f{ 551}{278737200}\chi_{10}\chi_{8}
             -\f{241907}{541865116800}\chi_{10}\chi_6\chi_2      \nn\\ 
         & &  +\f{62391997}{1357697236654080}\chi_{10}\chi_2^4
           -\f{697}{1982131200}\chi_8^2\chi_2                               
           -\f{23048029}{125712707097600}\chi_8\chi_6\chi_2^2 \nn\\
         & & -\f{143590607}{108615778932326400}\chi_8\chi_2^5
             -\f{1951}{6192744192}\chi_6^3
             +\f{113390999}{1939567480934400}\chi_6^2\chi_2^3 \nn\\
         & & -\f{516613213}{837893151763660800}\chi_6\chi_2^6
            +\f{809523655}{405406222549329641472}\chi_2^9\nn\\
         &=& -{\frac {221}{229360896}}\,{{ P_6}}^{3}
                +{\frac {3794551}{397367752320}}\,{{ P_2}}^{2}
                    { P_8}\,{ P_6}
           -{\frac {71315}{7224868224}}\,{ P_2}\,{ P_6}\,{ P_{10}}\nn\\
         & &  +{\frac {13525316017}{29663343763587072}}
             \,{{ P_2}}^{6}{ P_6}
                     -{\frac {55153997}{22888382533632}}\,{{ P_2}}^{3}
                  {{ P_6}}^{2}
                  -{\frac {473}{79639200}}\,{ P_2}\,{{ P_8}}^{2}\nn\\
         & &      +{\frac {119137}{681201861120}}\,{{ P_2}}^{3}{ P_{12}}
          +{\frac {453366913}{514988607006720}}\,{{ P_2}}^{4}{ P_{10}}
                 -{\frac {187565459}{205995442802688}}
                   \,{{ P_2}}^{5}{ P_8} \nn\\
         & &        -{\frac {703}{210247488}}\,{{ P_2}}^{2}{ P_{14}}
                         +{\frac {73}{9556704}}\,{ P_{12}}\,{ P_6}
                    +{\frac {157}{27873720}}\,{ P_{10}}\,{ P_8}\nn\\
         & &          -{\frac {142714197301}{6989762457747062784}}
                     \,{{ P_2}}^{9}+{\frac {1}{29496}}\,{ P_{18}}
        \nn\\
\eeqs  

\ni There are the following relations between $\chi$'s,
      
\beqs
\chi_4    &=& \f{\chi_2^2}{12} \nn\\
\chi_{16} &=& {\frac {13}{27}}\,{ \chi_6}\,{ \chi_{10}}
              +{\frac {13}{80}}\,{{ \chi_8}}^{2}
                +{\frac {590}{957}}\,{ \chi_2}\,{ \chi_{14}}
               -{\frac {8567}{31320}}\,{ \chi_2}\,{ \chi_8}\,{ \chi_6}\nn\\
          & &   -{\frac {15925}{103356}}\,{{ \chi_2}}^{2}{ \chi_{12}} 
            +{\frac {61607}{1691280}}\,{{ \chi_2}}^{2}{{ \chi_6}}^{2}
                +{\frac {5291}{338256}}\,{{ \chi_2}}^{3}{ \chi_{10}}\nn\\
          & &   +{\frac {7397}{10824192}}\,{{ \chi_2}}^{4}{ \chi_8}
               -{\frac {36127}{97417728}}\,{{ \chi_2}}^{5}{ \chi_6}
                  +{\frac {111449}{112225222656}}\,{{ \chi_2}}^{8}              
\eeqs  
    
We also have the relations between $\lambda$'s and FI parameters. We
only give the expressions for the first twenty eight weights out of
fifty six since the other twenty eight $\lambda$'s are minus the
weights given below. This reflects the fact that the {\bf 56} is a
real representation.

\begin{eqnarray}
& & v\cdot\lambda_1= -\f{3}{4}b_1-\f{1}{2}b_2-\f{1}{4}b_3
                                                +\f{1}{4}b_5  
                                            +\f{1}{2}b_6+\f{3}{4}b_7 \nn\\
& & v\cdot\lambda_2= -\f{3}{4}b_1
                                          -\f{1}{2}b_2-\f{1}{4}b_3
                                      +\f{1}{4}b_5+\f{1}{2}b_6-\f{1}{4}b_7\nn\\   & & v\cdot\lambda_3=-\f{3}{4}b_1-\f{1}{2}b_2-\f{1}{4}b_3+\f{1}{4}b_5 
                                            -\f{1}{2}b_6-\f{1}{4}b_7 \nn\\
& & v\cdot\lambda_4=-\f{3}{4}b_1-\f{1}{2}b_2
                                      -\f{1}{4}b_3-\f{3}{4}b_5-\f{1}{2}b_6
                                               -\f{1}{4}b_7 \nn\\            
& & v\cdot\lambda_5=-\f{1}{2}b_1+\f{1}{2}b_3+\f{1}{2}b_8  \nn\\
& & v\cdot\lambda_6=-\f{1}{2}b_1-\f{1}{2}b_3+\f{1}{2}b_8  \nn \\
& & v\cdot\lambda_7=-\f{1}{2}b_1+\f{1}{2}b_3-\f{1}{2}b_8  \nn \\
& & v\cdot\lambda_{8}=-\f{1}{2}b_1-b_2-\f{1}{2}b_3+\f{1}{2}b_8  \nn\\
& & v\cdot\lambda_{9}=-\f{1}{2}b_1-\f{1}{2}b_3-\f{1}{2}b_8 \nn\\
& & v\cdot\lambda_{10}=-\f{1}{2}b_1-b_2-\f{1}{2}b_3-\f{1}{2}b_8  \nn\\
& & v\cdot\lambda_{11}=-\f{1}{4}b_1+\f{1}{2}b_2+\f{1}{4}b_3+\f{3}{4}b_5
                                 +\f{1}{2}b_6+\f{1}{4}b_7  \nn\\
& & v\cdot\lambda_{12}=-\f{1}{4}b_1-\f{1}{2}b_2+\f{1}{4}b_3+\f{3}{4}b_5
                             +\f{1}{2}b_6+\f{1}{4}b_7 \nn\\
& & v\cdot\lambda_{13}=-\f{1}{4}b_1+\f{1}{2}b_2+\f{1}{4}b_3-\f{1}{4}b_5
                             +\f{1}{2}b_6+\f{1}{4}b_7 \nn\\
& & v\cdot\lambda_{14}=-\f{1}{4}b_1-\f{1}{2}b_2-\f{3}{4}b_3+\f{3}{4}b_5
                               +\f{1}{2}b_6+\f{1}{4}b_7 \nn\\
& & v\cdot\lambda_{15}=-\f{1}{4}b_1-\f{1}{2}b_2+\f{1}{4}b_3-\f{1}{4}b_5
                              +\f{1}{2}b_6+\f{1}{4}b_7 \nn\\
& & v\cdot\lambda_{16}=-\f{1}{4}b_1+\f{1}{2}b_2+\f{1}{4}b_3-\f{1}{4}b_5
                              -\f{1}{2}b_6+\f{1}{4}b_7 \nn\\       
& & v\cdot\lambda_{17}=-\f{1}{4}b_1-\f{1}{2}b_2-\f{3}{4}b_3-\f{1}{4}b_5
                              +\f{1}{2}b_6+\f{1}{4}b_7 \nn\\   
& & v\cdot\lambda_{18}=-\f{1}{4}b_1-\f{1}{2}b_2+\f{1}{4}b_3-\f{1}{4}b_5
                              -\f{1}{2}b_6+\f{1}{4}b_7 \nn\\   
& & v\cdot\lambda_{19}=-\f{1}{4}b_1+\f{1}{2}b_2+\f{1}{4}b_3-\f{1}{4}b_5
                               -\f{1}{2}b_6-\f{3}{4}b_7 \nn\\  
& & v\cdot\lambda_{20}=\f{1}{2}b_5+b_6+\f{1}{2}b_7+\f{1}{2}b_8 \nn\\  
& & v\cdot\lambda_{21}=-\f{1}{4}b_1-\f{1}{2}b_2-\f{3}{4}b_3-\f{1}{4}b_5
                               -\f{1}{2}b_6+\f{1}{4}b_7   \nn\\ 
& & v\cdot\lambda_{22}=-\f{1}{4}b_1-\f{1}{2}b_2+\f{1}{4}b_3-\f{1}{4}b_5
                               -\f{1}{2}b_6-\f{3}{4}b_7\nn\\   
& & v\cdot\lambda_{23}=\f{1}{2}b_5+\f{1}{2}b_7+\f{1}{2}b_8 \nn\\ 
& & v\cdot\lambda_{24}=\f{1}{2}b_5+b_6+\f{1}{2}b_7-\f{1}{2}b_8 \nn\\   
& & v\cdot\lambda_{25}=-\f{1}{4}b_1-\f{1}{2}b_2-\f{3}{4}b_3-\f{1}{4}b_5
                                -\f{1}{2}b_6-\f{3}{4}b_7 \nn\\    
& & v\cdot\lambda_{26}=-\f{1}{2}b_5+\f{1}{2}b_7+\f{1}{2}b_8 \nn\\
& & v\cdot\lambda_{27}=\f{1}{2}b_5-\f{1}{2}b_7+\f{1}{2}b_8 \nn\\  
& & v\cdot\lambda_{28}=\f{1}{2}b_5+\f{1}{2}b_7-\f{1}{2}b_8 \nn\\
\label{weights}
\end{eqnarray}

\section*{Appendix B}

To directly compare our curve with the one in \cite{MN2}, we should
 express our result in terms of Casimir invariants of the
 SO(12)$\times$SU(2) subgroup of E$_7$. More specifically, let us
 consider the subgroup we get by removing the simple root corresponding
 to $b_2$. Then the simple root corresponding to $b_1$ becomes the
 simple root of $SU(2)$ and the rest becomes associated with the roots
 of $SO(12)$. The mass parameters in \cite{MN2} can be thought of as an
 orthonormal basis for the root space. The standard way of choosing
 such a basis for $SO$ algebras would in our case correspond to
 \beqs
  b_3 &=& m_5 - m_6 ,\nonumber\\
  b_4 &=& m_4 - m_5 ,\nonumber\\
  b_5 &=& m_3 - m_4 ,\nonumber\\
  b_6 &=& m_2 - m_3 ,\\
  b_7 &=& m_1 - m_2 ,\nonumber\\
  b_8 &=& m_5 + m_6 ,\nonumber  
 \eeqs  
 and since $b_1$ is already orthogonal to everything else it is simply
 equal to the $SU(2)$ simple root
 \beqs
   b_1 &=& \phi  
 \eeqs
Similar relations were found in \cite{DHOO} for three dimensional theories.
 Inserting these expressions into our formulas we find that the curves
 are equal up to the following rescalings of the basic variables in   
 (\ref{ourcurve})
 \beqs
 X & \rightarrow & i \f{X}{8} \nn\\
 Y & \rightarrow & -\f{Y}{4} \nn\\ 
 Z & \rightarrow & -\f{Z}{2} \label{rescaling}
 \eeqs
 
 \ni which turns (\ref{ourcurve}) into
 
 \beqs
 X^2 &=& Y^3-\left[+2Z^3+8\alpha_1 Z-16\alpha_0\right]\,Y \nn\\
     & &  -\left[ 4\beta_4 Z^4-8\beta_3 Z^3
          +16\beta_2 Z^2 - 32\beta_1 Z + 64\beta_0\right],
 \eeqs
 which, after a shift in $Z$ (using the notation of \cite{MN2})
 \beqs
  Z \rightarrow Z+\f{1}{6}\left(\f{\tilde{T}_{2}^{2}}{12}+T_{4}\right),
 \eeqs
 becomes exactly the curve given in \cite{MN2}.

\end{document}